\documentstyle[preprint,aps,prl,tighten]{revtex}

\begin{document}


\draft

\title{ Electromagnetic Scattering from Relativistic Bound States}

\author{ N. K. Devine and S. J. Wallace}

\address{ Department of Physics and Center for Theoretical Physics \\
          University of Maryland, College Park, MD 20742 }

\date{\today}

\maketitle

\begin{abstract}
The quasipotential formalism for elastic scattering from relativistic bound
states is formulated based on the instant constraint in the Breit frame.
The quasipotential electromagnetic current
is derived from Mandelstam's five-point kernel and obeys a
two-body Ward identity.
Breit-frame wave functions are obtained directly by solving
integral equations with nonzero total
three-momentum, thus accomplishing a dynamical boost.
Calculations of electron-deuteron elastic form factors illustrate the
importance of the dynamical boost versus
kinematic boosts of the rest frame wave functions.
\end{abstract}
\pacs{25.30.Bf, 24.10.Jv, 11.10.Qr, 11.10.St}

\narrowtext


    In the study of relativistic bound states and scattering processes for
two particles, it is possible to perform a reduction from four
dimensions to three dimensions to obtain a quasipotential formalism
\cite{BbSLT,TjonQPReview,WallaceMan89,GrossTwoBody82} .
 The quasipotential is the kernel of the three-dimensional equation,
which in general may be defined covariantly.
A quasipotential reduction is commonly used
in studies of the nucleon-nucleon (NN) interaction because it
provides a covariant dynamics which is similar to the Schr\"{o}dinger
dynamics.
For NN scattering, one generally assumes a quasipotential in the form
of meson-exchange interactions with coupling constants selected to
provide a realistic description
of the NN scattering data and deuteron properties \cite{Gross92} .

 In this work, a quasipotential reduction procedure is applied to the
Mandelstam five-point function to derive a conserved current
operator consistent with the quasipotential wave functions.
The quasipotential constraint on the five-point kernel and initial and
final states must be consistent with conservation of the photon momentum, $q$.
In the Breit frame, where $q ^{0} = 0$, we find that the instant constraint
$p ^{0} = p ^{'0} = 0$ is consistent, where $p$ is the relative
momentum in the initial state and $p'$ is the relative momentum
in the final state.
Moreover, the instant constraint leads to a gauge invariant formalism which
is symmetric in its treatment of the particles.
With the instant constraint in the Breit frame for electromagnetic
matrix elements, the initial and final wave functions
must be calculated with total three momentum
${\bf{P}} = -{1\over 2} {\bf{q}}$ for the initial state and
${\bf{P}} =  {1\over 2} {\bf{q}}$ for the final state.
Only for ${\bf{q}} = 0$ are the usual rest frame wave functions used.
For ${\bf{q}} \neq 0$, the required wave functions may be thought of in terms
of a boost of the rest frame wave function.
In the instant form of relativistic quantum mechanics
one encounters a similar form of boost, and it must be dynamical
in the sense that the generator of boosts depends on the interaction.
In this paper, we formulate the dynamical boost within
the instant quasipotential formalism.  We present calculations of deuteron
electromagnetic form factors to demonstrate the feasibility of
calculating the required wave functions in the Breit frame and
we illustrate the importance of the dynamical
boost.

Formally, the quasipotential is defined such that the
results for two-body scattering based on the
Bethe-Salpeter equation are reproduced by use of the quasipotential and
a three-dimensional propagator.
Consider the Bethe-Salpeter t-matrix,
\begin{eqnarray}
\label{TMatrixEq}
T(p,q;P) &=& K^{BS}(p,q;P) + \int\frac{d^{4}p'}{(2\pi)^{4}} K^{BS}(p,p';P)
 \: G_{0}^{BS}(p';P) \: T(p',q;P),   
\end{eqnarray}
where $G_{0}^{BS}=i S_{F(1)}(\frac{1}{2}P+p')
i S_{F(2)}(\frac{1}{2}P-p')$ is the propagator for two free particles,
and $K^{BS}$ is the kernel consisting of irreducible graphs.
This equation may be abbreviated as $T^{BS}=K^{BS}+K^{BS} G^{BS} T^{BS}$,
with implied four-dimensional
integration.  Note that $P=p_{1}+p_{2}$ is the total momentum and
$p=\frac{1}{2} (p_{1}-p_{2})$ is the relative momentum.

A bound state with mass M gives rise to a pole
at $ P^{0} = \hat{P}^{0} \equiv \sqrt{M^{2}+{\bf P}^{2}}: $
\begin{equation}
\label{TMatrixPole}
  T(p,q;P) = \frac{-i}{2\hat{P}^{0}} \frac{\Gamma(p;\hat{P})
  {\overline\Gamma}(q;\hat{P}) }{ P^{0} - \hat{P}^{0} }  ,
\end{equation}
where ${\overline\Gamma} = \Gamma^{\dagger}\gamma^{0}_{1}\gamma^{0}_{2}$
and terms regular when $P^{0} = \hat{P} ^{0}$ are omitted.

The same T-matrix and bound state vertex function can be produced with
a quasipotential propagator, $G_{0}^{QP}(p,P) = i g_{0} 2\pi\delta(C(p))$,
where $C(p)=0$ is the constraint which reduces integrations from four to
three dimensions:
\begin{equation}
\label{e:T_QP}
  T = K^{QP} + K^{QP} G^{QP}_{0} T .
\end{equation}
The quasipotential kernel is defined by:
\begin{equation}
\label{e:K_QP}
  K^{QP} = K^{BS} + K^{BS} (G^{BS}_{0} - G^{QP}_{0}) K^{QP} .
\end{equation}
For the bound state vertex, one has a homogeneous equation,
found by substituting Eq.~(\ref{TMatrixPole}) into Eq.~(\ref{e:T_QP}) and
retaining pole terms:
\begin{equation}
\label{ConstraintChangeEq}
  \Gamma(p;\hat{P}) = \int\frac{d^{4}p'}{(2\pi)^{4}} K^{QP}(p,p';\hat{P})
  G^{QP}_{0}(p';\hat{P}) \Gamma(p';\hat{P}) .
\end{equation}

 The two most frequently used quasipotential constraints are the
one-particle-on-shell formalism developed by Gross and collaborators
\cite{GrossTwoBody82,Gross92}, based on
$C(p) \equiv (p_{2}^{0} - \epsilon_{2}) = 0$
($\epsilon_{2}\equiv \sqrt{m_{2}^{2}+{\bf p}_{2}^{2}}$),
and the equally-off-shell formalism in the center-of-mass frame developed by
Blankenbecler and Sugar and Logunov and Tavkhelidze \cite{BbSLT,TjonQPReview},
based on $C(p) \equiv (p_{1}^{2} - p_{2}^{2}) / 2\sqrt{P^{2}}
= p\cdot P / \sqrt{P^{2}} = 0$.
 When one particle is on mass shell, there is an inherent and inconvenient
asymmetry of the formalism, but this can be overcome \cite{Gross92} .
 However, a manifestly symmetric treatment is afforded by the
equally-off-shell formalism.
 In the center of mass frame, the equally-off-shell constraint is
equivalent to an instant formalism because the constraint causes
interactions to have zero time-component of momentum transfer.
 An extension of the equally-off-shell formalism to the full Dirac space
for two fermions has been developed by Mandelzweig and Wallace
\cite{WallaceMan89} by incorporating the iterative parts of
cross-box Feynman graphs, using a form of the eikonal approximation.

To obtain the quasipotential reduction for electromagnetic
interactions, we start from the Mandelstam formalism \cite{Mandelstam55}
for a five-point function, $T _{5}$, which has a photon of
momentum $q$ coupled in all possible ways to the two particles
and exchanged mesons.
The five-point function, $T_{5}$, may be expressed as follows,
\begin{equation}
\label{e:T5BS}
  T_{5} = (1 + T G^{BS}_{0}) K^{BS}_{5} (1 + G^{BS}_{0} T),
\end{equation}
where four-dimensional integrations are implied.
The irreducible five-point kernel, $K^{BS}_{5}$, is given by coupling
the photon to particles one and two (lowest order impulse contributions)
plus coupling the photon to all possible internal lines of the two-body
kernel $K ^{BS}$.
The five-point function can be written with a single quasipotential
constraint applied consistently before, inside, and after the
five-point kernel,
\begin{equation}
\label{e:T5QP}
  T_{5} = (1 + T G^{QP}_{0}) K^{QP}_{5} (1 + G^{QP}_{0} T).
\end{equation}
Equations~(\ref{e:T5BS}) and~(\ref{e:T5QP}) imply that,
\begin{eqnarray}
\label{e:K5QP}
  K^{QP}_{5} &=& [1 + K^{QP} (G^{BS}_{0}-G^{QP}_{0})] \: K^{BS}_{5} \:
 [1 +  (G^{BS}_{0}-G^{QP}_{0}) K^{QP}].
\end{eqnarray}
Unlike some previous formulations \cite{Faustov72,Jaus76,Jaus84},
Eqs.~(\ref{e:T5QP}) and~(\ref{e:K5QP}) use a single constraint consistent
with four-momentum conservation, and contain propagation
of negative- as well as positive-energy nucleon states.
To extract the electromagnetic matrix element for elastic
scattering from the bound state, one substitutes
Eq.~(\ref{TMatrixPole}) into
Eq.~(\ref{e:T5QP}).  The desired electromagnetic matrix element
is proportional to the residue of the double pole term in the resulting
expression,
   $\langle J^{\mu} \rangle = -\overline{\Gamma} G^{QP}_{0} K^{QP}_{5}
 G^{QP}_{0} \Gamma /2 \sqrt{P'^{0}P^{0}}$,
with implied four-dimensional integrations.

Instant constraints in the Breit frame may be expressed covariantly as
$p \cdot \widetilde{P} =  p' \cdot \widetilde{P} = 0$, where
$P$ and $P'$ are the initial and final total momenta, and
$\widetilde {P} = (P' + P) / \sqrt{(P' + P)^{2}} = (1,{\bf 0})$
in the Breit frame.
They imply that the wave
functions needed are a different slice of the four-dimensional
wave function at each different value of ${\bf{P}}$.
Other choices for the quasipotential constraint are either inconsistent
with momentum conservation or with particle-exchange symmetry.
For example, the instant constraint in the rest frame of the
initial and final bound states
is not consistent with conservation of four-momentum of the virtual
photon for electromagnetic interactions.
Momentum conservation requires $P' = P + q$ and $p' = p
\pm {1 \over 2}q$, depending on which particle absorbs the virtual
photon momentum, $q$.
Thus,
$p'\cdot P' = p\cdot P + p\cdot q \pm {1 \over 2} q \cdot (P+q)$,
which is not consistent with both $p'\cdot P' = 0$ and $p\cdot P = 0$ at
all values of relative momentum (which is integrated).

\begin{figure}
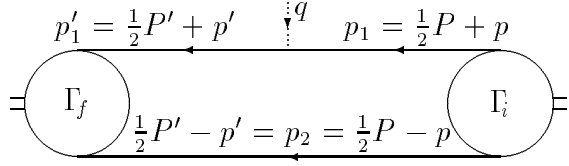

\caption{Lowest order impulse-approximation diagram with initial and final
vertex functions $\Gamma_{\!\!i}(p,P)$ and $\Gamma_{\!\!f}(p',P')$.
\label{f:ia}}
\end{figure}

An alternative is to place one particle on-shell \cite{Gross92},
say particle two in
Figure \ref{f:ia}.  This is consistent with momentum conservation
when the virtual photon is absorbed by particle 1, but not when
it is absorbed by particle 2.  Thus there is an inconsistency
with particle exchange symmetry, although for absorption by
the on-shell particle, it has been argued in Ref.~\cite{Gross&Riska}
that the difficulty can be overcome.

We conclude that conservation of four-momentum and particle exchange
symmetry suggest use of the instant constraint in the Breit frame.
We turn now to deriving the current operator and wave functions
based on Eq.~(\ref{e:K5QP}).

The impulse-approximation current operator for particle one may be
obtained by expanding $K_{5}^{QP}$ into diagrams, with the leading
order terms given by attaching the photon to particle one, together
with zero, one, or two exchanged bosons.
We have found that these leading diagrams yield an impulse current
operator which is consistent with the one-boson-exchange approximation
for the quasipotential, $K^{QP} \approx -i v^{OBE}$.
The homogeneous equation for the vertex, $(1 - K^{QP(OBE)} G^{QP}_{0})
\Gamma  =0$, is used to simplify the current with the result that
only the box and crossed-box five-point contributions
need be evaluated to obtain the current \cite{DevineThesis} .
The propagator that is consistent with this analysis is:
\begin{equation}
  G^{QP}_{0}(p;P) = i g_{0}({\bf p},P) (2\pi) \delta(p^{0}) ,
\end{equation}
where
\begin{equation}
    g_{0}({\bf p},P) = \sum_{\rho_{1},\rho_{2}} \frac{\Lambda_{1}^{\rho_{1}}
  ({\bf p}_{1}) \Lambda_{2}^{\rho_{2}}({\bf p}_{2}) }{
  (\rho_{1}+\rho_{2})(\epsilon_{D}/2) - \epsilon_{1} - \epsilon_{2}},
\end{equation}
$\rho_{i}=\pm$, $\rho_{i}\Lambda_{i}^{\rho_{i}}({\bf p}_{i})\gamma^{0}_{i}$
are projection operators for Dirac spinors with hermitian norm,
$\epsilon_{i}\equiv\sqrt{m^{2}_{N}+{\bf p}_{i}^{2}}$
(${\bf p}_{1,2}= {1\over 2}{\bf P}\pm {\bf p}$),
and $\epsilon_{D}\equiv \sqrt{M^{2}_{D}+{\bf P}^{2}} =P^{0}$.
Masses are $m_{N}$ (nucleon) and $M_{D}$ (deuteron).
This propagator is similar to the Dirac two-body propagator of
Ref.~\cite{WallaceMan89}, except for the use of the
instant constraint in the Breit frame instead of the equally-off-shell
constraint.

The five-point box and crossed-box diagrams are reduced to three
dimensional form in the Breit frame using a procedure similar to
Ref.~\cite{WallaceMan89}.
The crossed-box is evaluated using a form of the eikonal approximation,
and the relative energy of both
five-point diagrams is integrated with the retardation of the potentials
neglected.  The result is,
\begin{eqnarray}
\label{e:K5toJbc}
  K^{QP(IA1)}_{5}({\bf k}',P';{\bf k},P)
&=& \int\frac{d^{3}p}{(2\pi)^{3}} v^{OBE}({\bf k}',{\bf p}') \:
 g_{0}({\bf p}';P') \: \hat{J}_{IA}(1) \:
  g_{0}({\bf p};P) \: v^{OBE}({\bf p},{\bf k})
\end{eqnarray}
where the impulse-approximation current for particle one is,
\begin{equation}
\label{e:Jbc}
  \hat{J}^{\mu}_{IA}(1) \equiv \Gamma^{\mu}_{1} \gamma^{0}_{2}\hat{\rho}_{2}
\left( 1 + \frac{1-\hat{\rho}'_{1}\hat{\rho}_{1} }{2} \frac{2\epsilon_{2}
-\hat{\rho}_{2}\epsilon_{D} }{ \epsilon'_{1}+\epsilon_{1}} \right) ,
\end{equation}
$\hat{\rho}_{i} u^{\pm}(\pm {\bf p}_{i}) = \pm  u^{\pm}(\pm {\bf p}_{i})$,
and $\Gamma^{\mu}_{1}$ is the one-body electromagnetic operator.
The Breit frame total momenta are $P'=(\epsilon_{D},\frac{1}{2} {\bf q})$
and $P=(\epsilon_{D},-\frac{1}{2} {\bf q})$.

Using $K^{QP(IA1)}_{5}$  in
$\langle J^{\mu}\rangle = \overline{\Gamma} g_{0} K^{QP}_{5} g_{0} \Gamma
 /2\sqrt{P'^{0}P^{0}}$,
and $\psi \equiv g_{0} \Gamma = g_{0} v^{OBE} g_{0} \Gamma$,
the current matrix element for elastic scattering takes the form,
\begin{equation}
\label{e:Jem}
   \langle J^{\mu}_{IA}(1) \rangle = {1\over 2\epsilon_{D}}
 \int\frac{d^{3}p}{(2\pi)^{3}} \overline{\psi}({\bf p}+{1 \over 2}{\bf q};P')
 \hat{J}^{\mu}_{IA}(1) \psi({\bf p};P)
\end{equation}
A similar term is obtained for photon absorption by particle two.
The box-crossed current, $\hat{J}^{IA}(1)$, obeys an exact isoscalar,
two-body, Ward-Takahashi identity,
\begin{equation}
\label{e:tbWard}
  q \cdot \hat{J}_{IA}(1) = g^{-1}_{0}({\bf p}';P') - g^{-1}_{0}({\bf p};P) ,
\end{equation}
provided that the one-body current obeys the one-body WT identity,
$q \cdot \Gamma_{1} = S^{-1}_{(F)}(p'_{1})-S^{-1}_{(F)}(p_{1}) =
q \cdot \gamma_{1}$.  Thus working in the Breit frame and retaining the
five-point crossed-box allows for a gauge invariant quasipotential analysis.

The required wave functions may not in general be obtained by a kinematical
boost of rest frame wave functions.  Instead they must be calculated directly
in the Breit frame.  Thus the boost problem within the instant quasipotential
formalism involves a change of quasipotential constraint.
It is a straightforward matter to prove that the quasipotential must change
as follows,
\begin{equation}
\label{QPchange}
K^{QP2} = K^{QP1} + K^{QP1} (G^{QP1}_{0} - G^{QP2}_{0}) K^{QP2} ,
\end{equation}
where, for example, QP1 may refer to the instant
constraint in the rest frame, $C_{1}(p) = p\cdot P/\sqrt{P^{2}}$,
and QP2 may refer to the Breit frame instant constraint, $C_{2}(p) =
p\cdot\widetilde{P}$.
Equation~(\ref{QPchange}) defines formally how the interaction kernel
changes when a boost from the rest frame to the Breit frame is
performed, such that the same underlying covariant kernel
applies in the four-dimensional formalism.

Calculations of the instant wave functions in the Breit frame
are based on solving,
\begin{equation}
\label{e:homoeq}
  g^{-1}_{0}({\bf p}';P) \psi({\bf p}';P) =
  \int\frac{d^{3}p}{(2\pi)^{3}} v({\bf p}',{\bf p};P) \psi({\bf p};P) ,
\end{equation}
for ${\bf{P}}= \pm \bf{q}/2$.
An appropriate normalization condition is used.
A one-boson-exchange potential is used with scalar, pseudo-vector,
and vector mesons ($\sigma,\delta,\eta,\pi,\omega,\rho$)
to obtain deuteron wave functions using
modified Bonn B parameters \cite{DevineThesis,BonnABC}.

    The Breit frame total angular momentum operator is,
${\bf J} = {\bf J}_{1} + {\bf J}_{2} = \vec{{\cal L}} + {\bf S}$,
where $\vec{{\cal L}} = {\bf l} + {\bf L}$,
${\bf l}={\bf r} \times {\bf p}$, ${\bf L}= {\bf R} \times {\bf P}$, and
${\bf S} = \frac{1}{2} (\vec{\sigma}_{1}+\vec{\sigma}_{2})$.
Because ${\bf J}$ and $J^{z}$ commute with $g^{-1}_{0}$ and $v$,
solutions of the homogeneous equation~(\ref{e:homoeq}) are eigenfunctions
of ${\bf J}$ and $J^{z}$.
However, ${\bf l}$, ${\bf L}$,  and ${\bf S}$ separately do not commute with
$g^{-1}_{0}$.  To proceed, we define sixteen Dirac plane-wave basis functions,
\begin{equation}
\label{pwbasis}
  \chi^{\rho_{1},\rho_{2}}_{s_{1},s_{2};M_{J}}({\bf p},{\bf P}) \equiv
 u^{\rho_{1}}_{1}(\rho_{1} {\bf p}_{1}) u^{\rho_{2}}_{2}(\rho_{2} {\bf p}_{2})
 {\cal Y}^{M_{J}}_{s_{1},s_{2}}(\phi),
\end{equation}
where
\begin{equation}
    {\cal Y}^{M_{J}}_{s_{1},s_{2}}(\phi) =
  e^{i(M_{J}-s_{1}-s_{2})\phi} | s_{1} \rangle | s_{2} \rangle,
\end{equation}
and $s_{i}=\pm \frac{1}{2}$ and $J^{z} {\cal Y}^{M_{J}}
= M_{J} {\cal Y}^{M_{J}}$.

Because the usual partial-wave analysis is inapplicable,
the homogeneous equation is solved in three dimensions using the basis
functions (\ref{pwbasis}) with only $\phi$ integrations carried out
analytically.  Radial and polar angle integrations are performed numerically.
The homogeneous equation is solved for $M_{J}=0$ at fixed values
of total momentum using the Malfliet-Tjon iteration procedure \cite{MalfTjon}.
Wave functions with polarization states $M_{J}=\pm 1$ are obtained from
the $M_{J}=0$ state by using the raising and lowering operator,
 $\psi^{M_{J}\pm 1} = \sqrt{(J+M_{J})(J-M_{J}+1)} J^{\pm} \psi^{M_{J}}$.

\begin{figure}
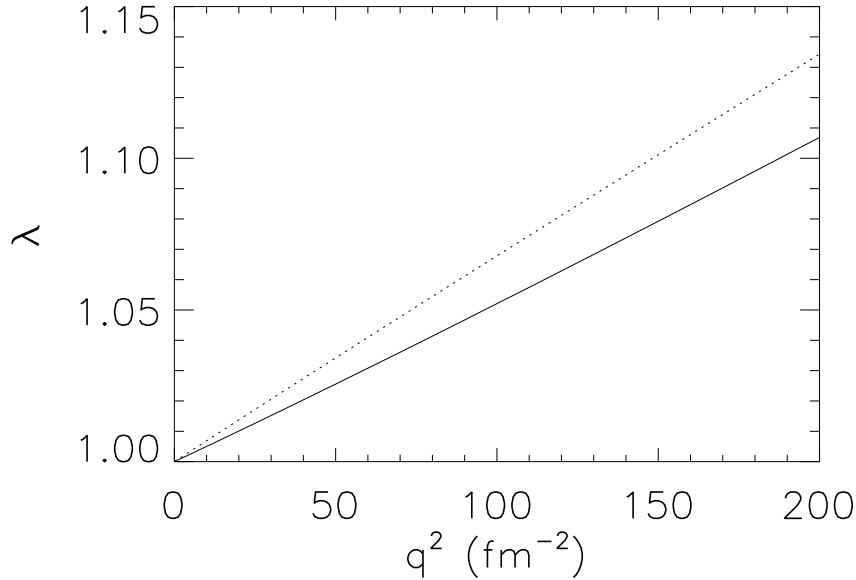

\caption{The scaling of the potential, $\hat{V}(p'-p,{\bf P}=
\pm {1\over 2}{\bf q})= \hat{V}(p'-p) / \lambda ({\bf q}^{2})$,
that produces constant deuteron mass,
$M_{D}= (\epsilon_{D}^{2}-{\bf P}^{2})^{1/2} = 2m_{N}-2.22464MeV$: full
propagator (solid), $++$ states only (dotted).
\label{f:lambdaVSqsq} }
\end{figure}

Variation of the quasipotential with total momentum is approximated in a
very simple manner, $\hat{V}({\bf p}'-{\bf p},{\bf P}=\pm {1\over 2}{\bf q})
= \hat{V}({\bf p}'-{\bf p}) / \lambda ({\bf q}^{2})$
where $\lambda ({\bf q}^{2})$ is fit to produce the correct deuteron total
energy, $\epsilon_{D}= (M_{D}^{2}+{\bf q}^{2}/4)^{1/2}$.
Figure~\ref{f:lambdaVSqsq} shows that the required change of the
potential is modest, with $\lambda$ varying linearly
over a wide range of $q^{2}$ values.
When $\lambda ({\bf q}^{2})=1$ is used, the potential is too attractive and
the binding energy of the deuteron increases from
$2m_{N}-M_{D} \approx 2.2MeV$ at ${\bf q}^{2}=0$ to
$2m_{N}-M_{D} \approx 4.2MeV$ at ${\bf q}^{2}=200fm^{-2}$.
The difference in the deuteron form factors produced by setting
$\lambda ({\bf q}^{2})=1$ is minor in comparison with the ambiguity in the
boost of rest frame wave functions.
Note that with the instant constraint $v^{(OBE)}$ as well as $g_{0}$
and $J_{IA}$ are non-singular.

\begin{figure}
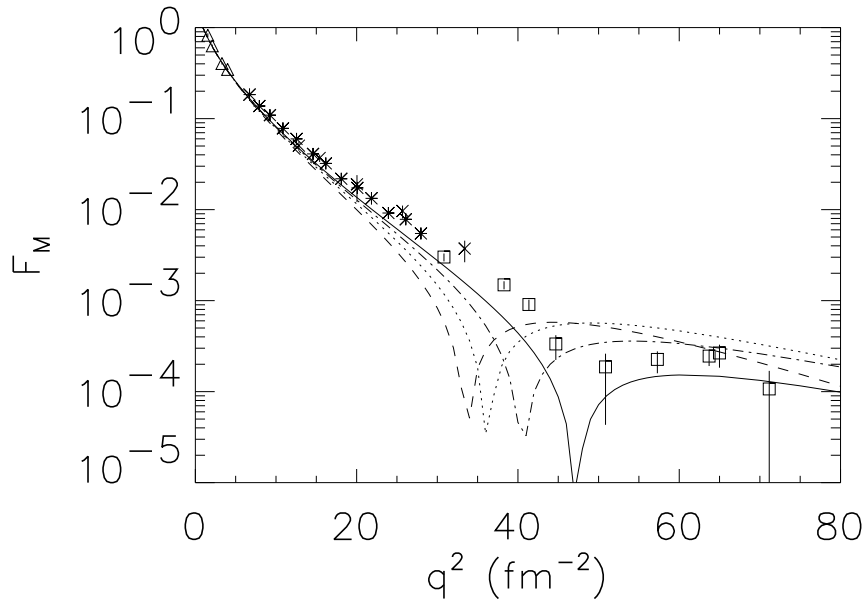

\caption{Elastic e-d magnetic form factor: consistent calculation with
IA+MEC (solid).  Impulse approximation only: consistent calculation
(dotted), boost approximations with $p^{0}(Breit)=0$ (dashed) and
$p^{0}(cm)=0$ (dash-dotted). See text.
\label{f:fmVSqsq} }
\end{figure}

Results for the deuteron magnetic form factor are shown in
Figure~\ref{f:fmVSqsq}.  The solid line result includes the impulse
approximation plus $\rho\pi\gamma$, $\omega\sigma\gamma$, and
$\omega\eta\gamma$ meson-exchange-currents,
 calculated
with the instant wave functions and current operators.  We use the
same meson-exchange-current operators as Hummel and Tjon
\cite{DevineThesis,HummelMEC,HummelThesis}
with $g_{\rho\pi\gamma}=.563$, $g_{\omega\sigma\gamma}=-.4$,
$g_{\omega\eta\gamma}=-.206$.
 To illustrate the importance of the dynamical boost, we compare the
impulse approximation contributions based on solving Eq.~(\ref{e:homoeq})
(dotted line) and two approximations
based on using the rest frame wave functions
with a kinematical boost as follows,
\begin{equation}
   \psi(p,P)=  \Lambda_{1}({\cal L}) \Lambda_{2}({\cal L})
  \psi({\bf p}_{rest},P_{rest}),
\end{equation}
where $P_{rest} = (m_{D},{\bf 0})$, $P = (\epsilon_{D},\pm {\bf q}/2)$, and
${\cal L} P_{rest} = P$.
The Lorentz transform of the relative momenta, ${\cal L} p_{rest} = p$, is
ambiguous since it is not possible to simultaneously satisfy
both constraints: ($p^{0}=0$) and ($p^{0}_{rest}=0$).
In Figure~\ref{f:fmVSqsq}, the dashed line is the result of
satisfying ($p^{0}=0$), while the dash-dotted line is the result of
satisfying ($p^{0}_{rest}=0$).

The instant quasipotential formalism in the Breit frame provides in principal
a solution to the long-standing problem of how to boost three-dimensional wave
functions with the instant constraint.
The significant differences in the results based on  instant wave
functions calculated directly in the Breit frame and the two approximations
to them suggest the importance of the formalism of this paper.



   We wish to thank Dr.\ A.\ Delfino for helpful discussions on the
Malfliet-Tjon procedure.  Support for this work by the U.S. Department of
Energy under grant DE-FG02-93ER-40762 is gratefully acknowledged.


\end{document}